\documentclass[twocolumn,superscriptaddress,showpacs]{revtex4}

\usepackage[english]{babel}
\usepackage{epsfig}
\usepackage{amsmath}
\usepackage{graphicx}

\newcommand{\bs}{\boldsymbol}

\pacs{47.10.-g,47.27.-i,47.27.De}

\begin{document}

\title{Lagrangian Investigation of Two-Dimensional Decaying Turbulence}

\author{Michael Wilczek}
\email{mwilczek@uni-muenster.de}
\author{Oliver Kamps}
\author{Rudolf Friedrich}
\affiliation{Institute for Theoretical Physics, University of M\"unster, Wilhelm-Klemm-Str. 9, 48149 M\"unster, Germany}

\begin{abstract}
We present a numerical investigation of two-dimensional decaying turbulence in the Lagrangian framework. Focusing on single particle statistics, we investigate Lagrangian trajectories in a freely evolving turbulent velocity field. The dynamical evolution of the tracer particles is strongly dominated by the emergence and evolution of coherent structures. For a statistical analysis we focus on the Lagrangian acceleration as a central quantity. For more geometrical aspects we investigate the curvature along the trajectories. We find strong signatures for self-similar universal behavior.
\end{abstract}

\maketitle

\textit{\bf Introduction}
Beyond its importance for geophysical and astrophysical applications two-dimensional turbulence often serves as a paradigm of a complex, self-organizing system. Despite its spatio-temporal complexity freely decaying two-dimensional turbulence exhibits a stunning degree of coherence. The dynamical evolution of the turbulent vorticity field is dominated by the emergence, coalescence and nonlinear interaction of circular and spiral-like vortices (see e.g. \cite{tabeling02prp} and references therein).\par 
During the decay process these coherent structures tend to organize in a self-similar way \cite{benzi88jpa,laval01pre,carnevale91prl,weiss93pfa}, for example the temporal evolution of their density, the mean absolute value of the circulation and the mean vortex radius follow power-laws \cite{carnevale91prl}, indicating that the decay process might display some degree of universality.\par
The Lagrangian frame of reference offers a natural access to the description of turbulent flows. However, results on the Lagrangian description of two-dimensional turbulence are sparse. Hence we are led to study the evolution of tracer particles in the two-dimensional decaying system. We find, as expected, that their evolution is strongly dominated by the coherent vortices.\par 
Regarding a statistical analysis we focus on the acceleration as well as on the curvature along the trajectories. We find strong evidence for a self-similar temporal evolution of some of the quantities. \par
The system under consideration is governed by the nonlinear effects of vortex interaction and merging as well as by viscous forces. Thus please note that our results cannot be referred to in the framework of the universal decay theory proposed by Carnevale et al. \cite{carnevale91prl,weiss93pfa} as we operate in a different parameter range, where viscosity cannot be neglected.\par 
The remainder of this article is structured as follows. After summarizing details on numerical issues and simulation parameters we procede to a qualitative discussion of trajectories. We then present statistical results of the Lagrangian acceleration and curvature.\par

\textit{\bf Numerics and Simulation Details}
\begin{figure}
    \includegraphics[width=0.4\textwidth]{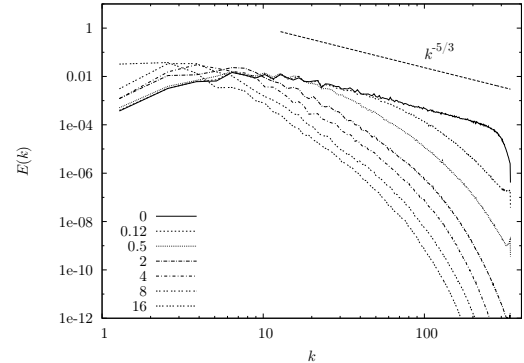}\\
    \caption{Temporal evolution of the energy spectrum for $t \in [0,16]$. High modes get damped out while the energy contained in the low modes increases.}
    \label{fig:energyspectrum}
\end{figure}
\begin{figure}
    \includegraphics[width=0.45\textwidth]{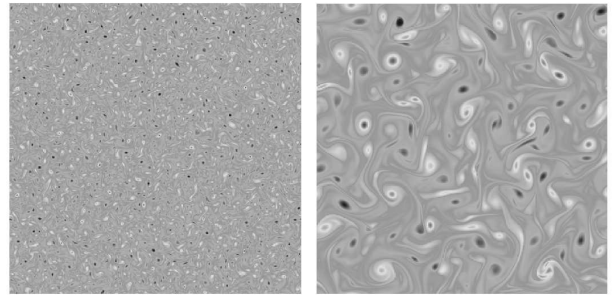}\\
    \caption{Vorticity field for $t=0.3$ and $t=4$. The number of vortices decreases due to vortex coalescence and destruction.}
    \label{fig:vorticity}
\end{figure}
We solve the two-dimensional vorticity equation
\begin{equation}
  \frac{\partial \omega}{\partial t}(\bs x,t)+(\bs u(\bs x,t)\cdot\nabla)\omega(\bs x,t)=\nu\Delta\omega(\bs x,t)
\end{equation}
by means of a standard pseudospectral code on a $1024^2$ grid with periodic boundary conditions. Table \ref{tab:simpara} sums up the major simulation parameters. We suppress aliasing errors by a spherical mode truncation according to Orszag's famous two thirds rule. To obtain reliable statistical results we immerse approximately 16000 tracer particles into the flow and follow them throughout the decay process. The Eulerian fields are interpolated by a bicubic scheme. The acceleration is explicitly calculated each timestep by evaluating the right-hand side of the Navier-Stokes equation. The resulting field is then interpolated at the particle positions.\par
  
In order to increase statistical quality we additionally perform an ensemble average over forty independent realizations of the flow (about 19 gigabytes of Lagrangian tracer data, ${\cal O}(1600)$ hours of cpu time). The initial conditions are obtained from a forced turbulence simulation that has reached statistical equilibrium in the following manner. A master field from such a forced simulation is taken and forty copies are made. For each copy a forced simulation with a duration of about five large eddy turnover times is performed with a differently initialized random number generator for the forcing (see \cite{kamps07unp} for details on the forcing). We thereby gain forty statistically independent initial conditions for the subsequent decaying turbulence runs. We would like to stress that by the large amount of ensemble realizations we achieve an extraordinarily high statistical quality. One additional run with doubled total simulation time was performed for checking and visualization purposes. For a more detailed description of the numerical issues see reference \cite{kamps07unp}.\par 
The initial conditions exhibit a clear inertial range according to Kolmogorov's predictions as depicted in figure \ref{fig:energyspectrum} (here, no ensemble average was performed). The time-resolved energy spectrum reveals that the initial inertial range scaling is quickly destroyed as the high modes get damped out. Simultaneously energy is transported into the low modes by nonlinear mode interaction of the inverse energy cascade. In physical space this is mirrored by the decay of small-scaled structures and the emergence of bigger coherent vortices as depicted in figure \ref{fig:vorticity}.\par

\begin{table}
  \begin{tabular}{cccccccc}
     $N^2$    & $L$    & $\nu$    & $Re_{i}$ & $dt$     & $T$  & $N_{\mathrm{tra}}$ & $N_{\mathrm{re}}$ \\
     \hline
     $1024^2$ & $2\pi$ & $0.0004$ &  $12500$ & $0.0005$ & $25$ & $16384$            & $40$ \\
   \end{tabular}
  \caption{Major simulation parameters. Number of grid points $N^2$, box length $L$, viscosity $\nu$, initial Reynolds number $Re_i=\frac{2\pi \, u_{\mathrm{rms}}}{\nu}$, timestep $dt$, total simulation time $T$, number of tracer particles per realization $N_{\mathrm{tra}}$ and total number of realizations $N_{\mathrm{re}}$. }
  \label{tab:simpara}
\end{table}

\textit{\bf Trajectories}
\begin{figure}
    \includegraphics[width=0.35\textwidth]{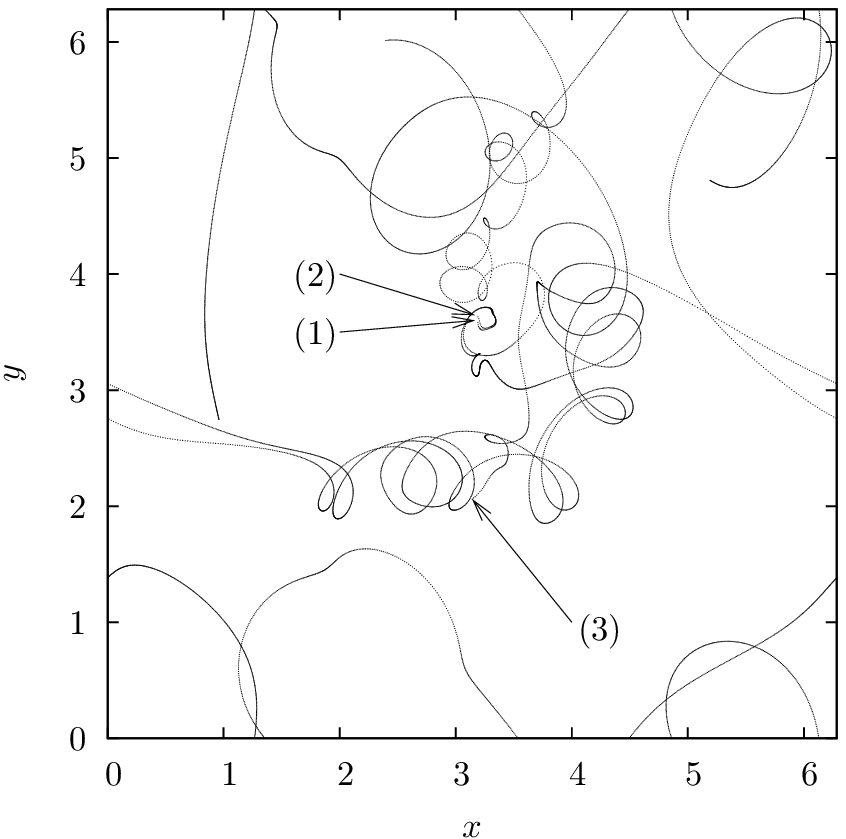}\\
    \caption{Typical particle paths in decaying turbulence. The spiraling motion indicates vortex trapping events. The numbers denote the initial position of the three particles.}
    \label{fig:tra1}
\end{figure}
\begin{figure}
    \includegraphics[width=0.45\textwidth]{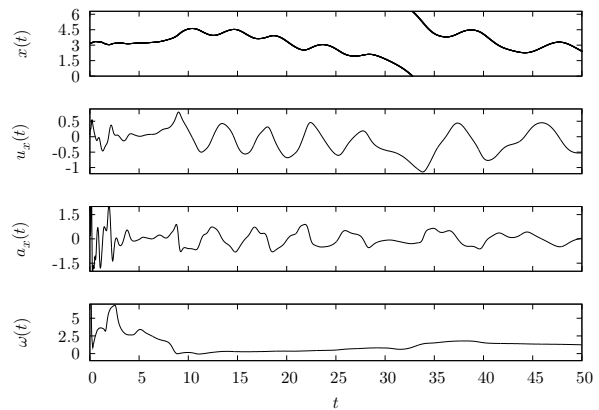}\\
    \caption{Time-resolved trajectory in decaying turbulence. Vortex trapping results in an oscillating behavior of the components of position, velocity and acceleration.}
    \label{fig:tra2}
\end{figure}
Turning now to a Lagrangian description of the turbulent field, figure \ref{fig:tra1} shows some sample trajectories. As characteristic for a turbulent chaotic system two initially neighboring trajectories ((1) and (2)) separate quickly as they get caught into the velocity field of different vortices. The spiraling motion of the particles indicates that vortex trapping is generic for particles in two-dimensional decaying turbulence. As the vortices merge into bigger structures the radius of the spiral motion increases on average over time. We postpone a more quantitative discussion of this point to the following section. \par 
While the chaotic advection can significantly increase the pair separation also the opposite takes places. For example trajectory (1) and trajectory (3) subsequently enter the velocity fields of the same vortex leading to a sticking effect; the vortex trapping event causes these two particles to stay close for a comparably long time although they started from far separated places in the turbulent field. \par
Figure \ref{fig:tra2} shows the time-resolved trajectory (1). After the initial emergence of medium-scaled vortices from the rather rough and unstructured initial vorticity all the components of space, velocity and acceleration show the signatures of vortex trapping by their oscillating behavior. The strength and also the frequency of the oscillations decrease throughout the decay process as the coherent structures increase in size and the velocity field gets smoother and damped because of viscous effects. All these observations show that Lagrangian particle dynamics and hence statistics is strongly dominated by the motion in single coherent structures.\par\bigskip 

\textit{\bf Statistical Results}
After this rather qualitative discussion let us now turn to the central statistical observable of this work. The Lagrangian acceleration of a particle initially located at $\bs y$ is defined by
\begin{equation}
  \bs a(t,\bs y)=\ddot{ \bs X}(t,\bs y)=\left[-\nabla p(\bs x,t)+\nu \Delta \bs u(\bs x,t) \right]_{\bs x=\bs X(t,\bs y)}.
\end{equation}
For a more detailed dynamical description this acceleration can be decomposed into a acceleration parallel to the current velocity and a acceleration perpendicular to the current velocity. With $\bs u_{\parallel}=\bs u/|\bs u|$ and $\bs u_{\perp}=(-u_y,u_x)/|\bs u|$ the decomposition reads
\begin{equation}
  \bs a(t,\bs y)=a_{\parallel}(t,\bs y) \, \bs u_{\parallel} +a_{\perp}(t,\bs y) \, \bs u_{\perp},
\end{equation}
thereby defining
\begin{equation}
  a_{\parallel}=\bs a\cdot \bs u_{\parallel} \quad \mathrm{and} \quad a_{\perp}=\bs a\cdot \bs u_{\perp}.
\end{equation}
\begin{figure}
    \includegraphics[width=0.45\textwidth]{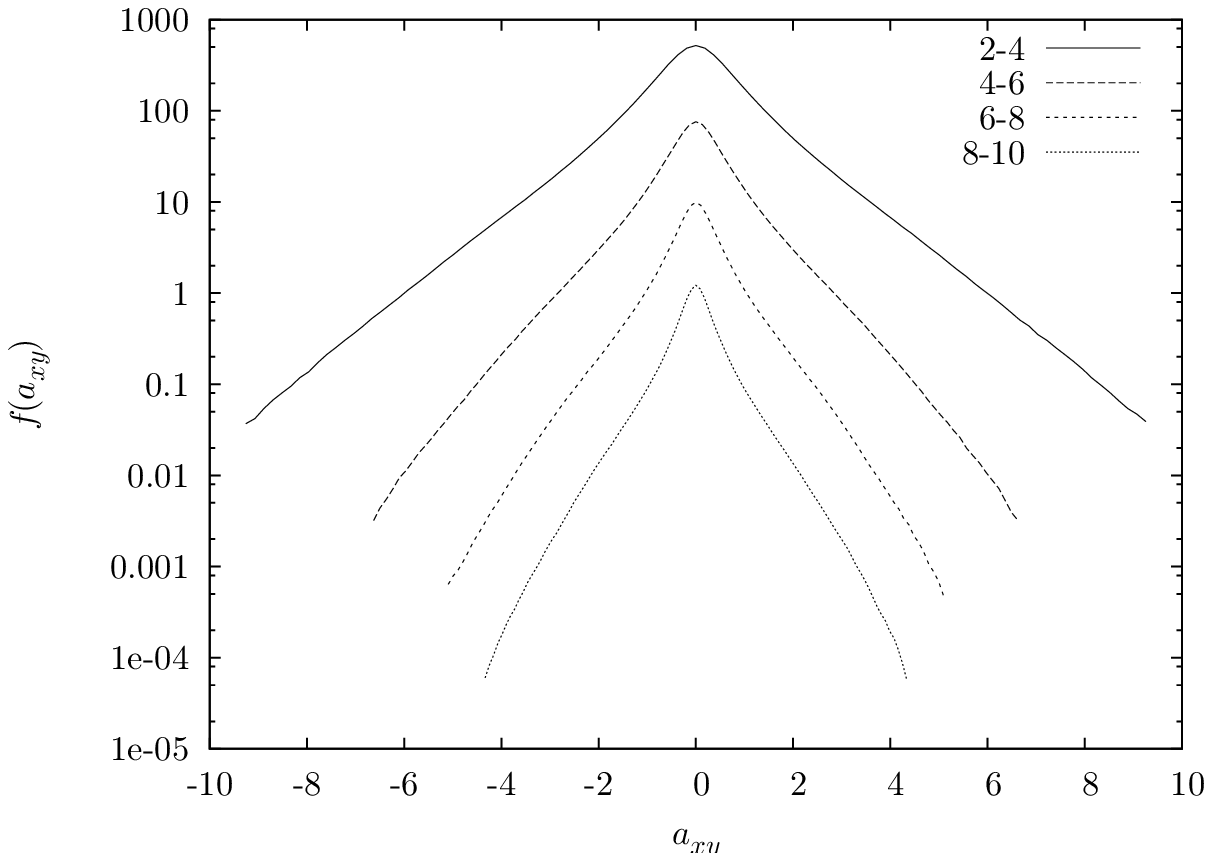}\\
    \includegraphics[width=0.45\textwidth]{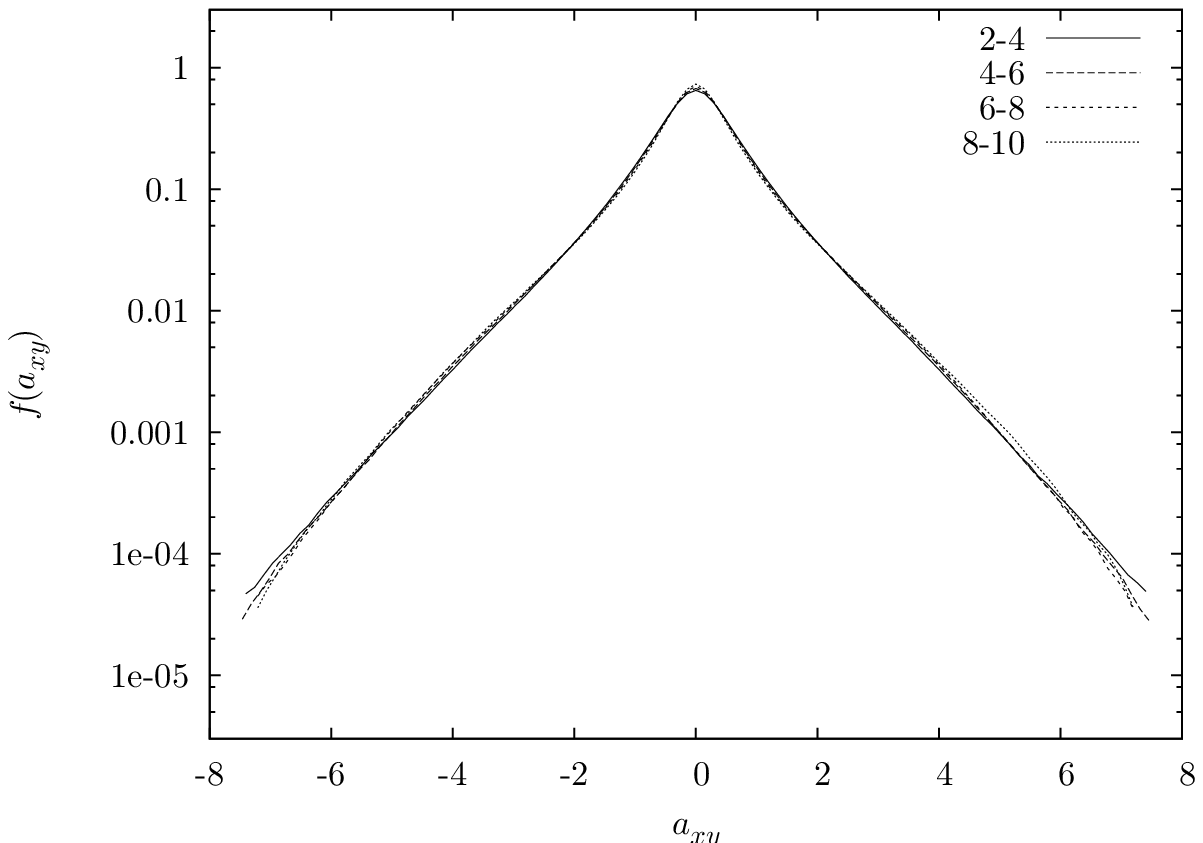}\\
    \caption{Time-resolved acceleration pdf. The upper figure shows the acceleration pdf's for different times vertically shifted. The lower figure shows the same pdf's normalized by their standard deviation resulting in a collapse of the curves. Time-windows over which averaging is performed are indicated.}
    \label{fig:fax}
\end{figure}
\begin{figure}
    \includegraphics[width=0.45\textwidth]{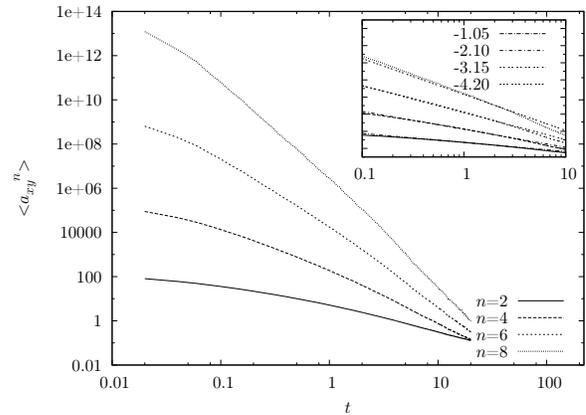}\\
    \caption{Time-resolved moments of the acceleration pdf. The moments clearly show power-law behavior. The inset shows a comparison with power-law functions. The algebraic exponents are displayed.}
    \label{fig:momax}
\end{figure}
Let us start with the time-resolved pdf $f(a_{xy})$ of a single component of the acceleration, which is depicted in figure \ref{fig:fax}. Owing to the statistical isotropy of the flow we improve our statistics by averaging over both spatial directions.
The time-resolved acceleration pdf's clearly exhibit large deviations from a Gaussian distribution as expected from the intermittent nature of turbulent acceleration. They exhibit almost exponential tails. However, when normalized by their standard deviation all of the curves collapse over the selected time interval thereby indicating a universal temporal evolution of the pdf. This view is supported by figure \ref{fig:momax} which shows the time-resolved moments of the acceleration pdf. By inspection of this figure the scaling regimes roughly can be estimated as the time interval for $t \in [0.1,10]$. Note that this choice is rather conservative. The size of this time interval will surely depend on physical parameters like viscosity or initial Reynolds number. Additionally statistical issues might matter; as the turbulence decays the number of vortices rapidly drops. Consequently there are fewer and fewer coherent structures contributing to the overall statistics resulting in a decreasing statistical quality.\par
Figure \ref{fig:momax} also shows a comparison with power-law functions $\langle a_{xy}^n \rangle\sim t^{\zeta_n}$ with the exponents $\zeta_n$. The numerical values roughly suggest $\zeta_{2n} \sim n\zeta_2$ for the selected moments as necessarily required for a self-similar evolution of the pdf's. Please note that we do not claim these exponents to be universal for all turbulent flows as they should dependend on the Reynolds number, the type of numerical viscosity applied in the simulation and on the exact initial condition. For a detailed investigation of the impact of different initial conditions on the decay process see \cite{bokhoven07pof}.\par
\begin{figure}
    \includegraphics[width=0.45\textwidth]{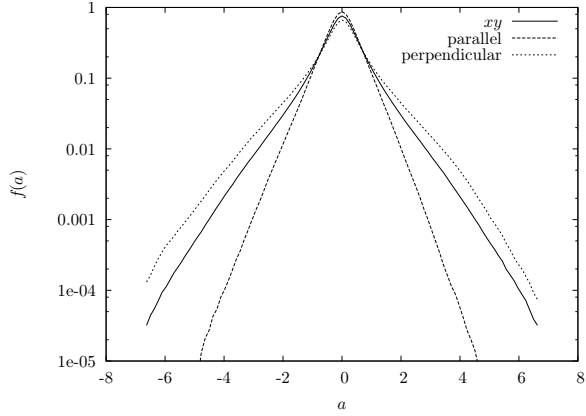}\\
    \caption{Pdf's of the acceleration components $a_{xy}$, $a_{\parallel}$ and $a_{\perp}$ for the time-window 4-6. In comparison with $a_{xy}$ the perpendicular component is more intermittent while the parallel component is less intermittent.}
    \label{fig:faparperp}
\end{figure}

 
We now proceed to a discussion of the parallel and perpendicular component of the acceleration. Figure \ref{fig:faparperp} shows a comparison of the pdf of the acceleration components $a_{xy}$, $a_{\parallel}$ and $a_{\perp}$. One can see that the perpendicular component is slightly more intermittent than the $xy$-averaged component. However, the parallel component is clearly less intermittent. This difference remains when the pdf's are rescaled to their standard deviation (not depicted) indicating that the three components have a fundamentally differing functional form. The physical interpretation of these results is quite straightforward. The fact that the perpendicular component of the acceleration is more intermittent than the parallel one shows that the spiraling motion in the vortices is a major contributor to the intermittent nature of the turbulent acceleration. Hence this is statistical evidence for the dynamical importance of vortices regarding Lagrangian dynamics.\par 
\begin{figure}
    \includegraphics[width=0.45\textwidth]{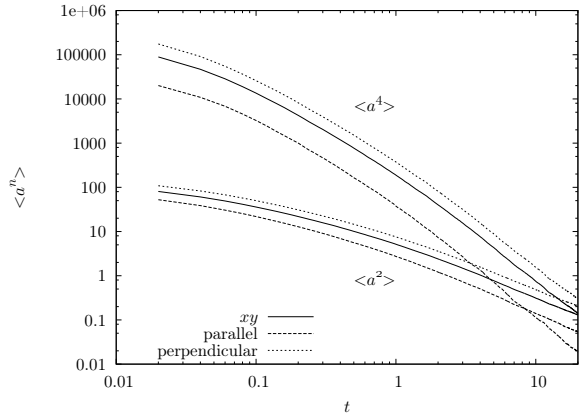}\\
    \caption{Temporal scaling for the second and fourth moment of the acceleration components $a_{xy}$, $a_{\parallel}$ and $a_{\perp}$. While the scaling exponents seem to be almost identical the moments differ by a multiplicative factor.}
    \label{fig:momparper}
\end{figure}
We checked that also the pdf's of the components $a_{\parallel}$ and $a_{\perp}$ collapse over the selected interval when normalized by their standard deviation (not depicted) and hence also the moments show a power-law behavior. This is exemplified in figure \ref{fig:momparper}, where the second and fourth moment of the corresponding pdf's are shown. This inspection reveals that the moments show nearly identical temporal scaling behavior, i.e. $\langle a_{xy}^n \rangle = c_{xy} \, t^{\zeta_n}$, $\langle a_{\parallel}^n \rangle = c_{\parallel} \, t^{\zeta_n}$ and $\langle a_{\perp}^n \rangle = c_{\perp} \, t^{\zeta_n}$ with identical $\zeta_n$, but differ by their prefactors $c_{xy}$, $c_{\parallel}$ and $c_{\perp}$. That means the differing functional form can be traced back to a differing weighting for each moment. The emerging picture is quite interesting: while the functional form of each of these pdf's seems to be determined by an initial weighting factor for each moment, their temporal behavior is universal. \par 
To close this section let us turn to more geometrical properties of the trajectories. We investigate the curvature of the trajectories defined by
\begin{equation}
   \kappa(t,\bs y)=\frac{|a_{\perp}(t,\bs y)|}{\bs u^2(t,\bs y)}.
\end{equation}
\begin{figure}
    \includegraphics[width=0.45\textwidth]{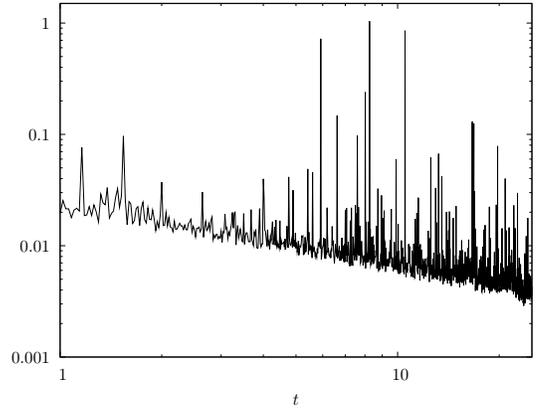}\\
    \caption{Log-log plot of the mean curvature as a function of time. The increasing spiral radii of the trajectories result in a decaying curvature.}
    \label{fig:curvature}
\end{figure}
\begin{figure}
    \includegraphics[width=0.45\textwidth]{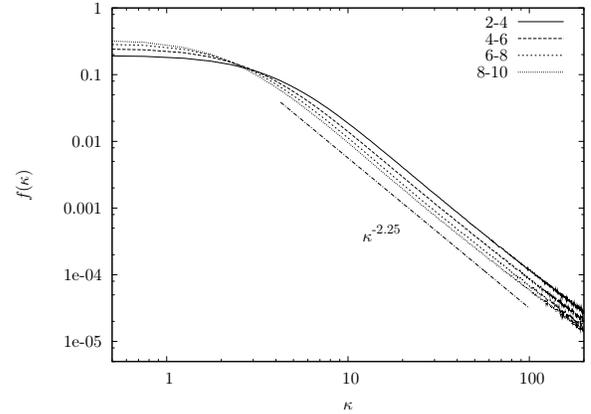}\\
    \caption{Log-log-plot of the time-resolved curvature pdf. The pdf's show an algebraic decay with an exponent close to $-2.25$.}
    \label{fig:pdf_curvature}
\end{figure}
Figure \ref{fig:curvature} shows the mean curvature, averaged over all trajectories. Although this quantity is extremely fluctuating, that is because of the $1/\bs u^2$ dependence, the average trend of a decreasing curvature is clearly visible. As the curvature is proportional to the inverse of the spiral radius, the emergence of bigger and bigger coherent structures therefore on average leads to increasing spiral radii of the trajectories. This is the statistical approval for the qualitative discussion of the trajectories in the preceding section. The time-resolved pdf of the curvature contains even more information and is shown in figure \ref{fig:pdf_curvature}. The time-resolved pdf's are fairly similar to each other. However, the probability for a high curvature (a small spiral radius) decreases over time, as expected for coalescing and growing vortices. Interestingly, all of the pdf's show an extraordinarily clear algebraic decay with an exponent close to $-2.25$.\par

\textit{\bf Summary}
To sum up, we presented a detailed investigation of two-dimensional decaying turbulence in the Lagrangian frame of reference. Our results reveal that the Lagrangian dynamics in this system is strongly influenced by long periods of trapping in coherent structures. The spiraling motion of the particles is a consequence.\par
We focused on the Lagrangian acceleration as a central statistical quantity finding that the turbulent fields tend to organize in a self-similar way. Hence we are able to identify a time interval where a self-similar scaling of the acceleration pdf's and corresponding moments holds. A decomposition of the acceleration into components parallel and perpendicular to the velocity reveals that the intermittent nature of the acceleration is partly due to the centripetal accelerations caused by the coherent structures. Closely connected to the acceleration is the curvature of the trajectory, adding a more geometrical point of view. We found the mean curvature decaying algebraically in time, consistent with the qualitative picture drawn in the first sections. The time-resolved pdf's of the curvature show a clear power-law behavior.\par 
Our investigation gives evidence that two-dimensional decaying Lagrangian turbulence is dominated by self-similar scaling laws, even when viscosity is not negligible.

\end{document}